\documentclass{PoS}
\usepackage{amsmath,amssymb}
\usepackage{mathtools}
\usepackage{amsfonts}
\usepackage{pstricks}
\usepackage{amsthm}
\usepackage{mathrsfs}

\title{Dark matter in GUT inspired $Z^\prime$ portal scenarios}

\ShortTitle{Dark matter in GUT inspired $Z^\prime$ portal scenarios}

\author{\speaker{Mathias Pierre}\thanks{Preprint number : LPT-Orsay-16-55}\\
       Laboratoire de Physique Th\'eorique, CNRS, Univ. Paris-Sud, Universit\'e Paris-Saclay, 91405 Orsay, France \\
        E-mail: \email{mathias.pierre@th.u-psud.fr}}

\abstract{We consider simple dark matter extensions of the standard model in $Z^\prime$ portal scenarios inspired by grand unification theory constructions and we study the phenomenology of such models by considering Spin Dependent and Spin Independent direct detection constraints, confronting the canonical thermally produced dark matter  scenario and we show that in this simple framework the combination and the complementary of these constraints is a powerful tool to derive stringent bounds and to reduce the viable parameter space of the model.
}

\FullConference{Proceedings of the Corfu Summer Institute 2015 "School and Workshops on Elementary Particle Physics and Gravity"\\
		1-27 September 2015\\
		Corfu, Greece}

\begin{document}

\section{Introduction}

While the existence of Dark Matter (DM) seems to be well established from the observation of large scale structures, N-body simulations and measurements of the anisostropies of the Cosmic Microwave Background~\cite{planck_collaboration_planck_2015}, its exact nature and properties remain misunderstood and still not described by the Standard Model (SM). Constraints have been set on the dark matter couplings to SM fields in universal frameworks~\cite{del_nobile_tools_2013} with effective field theory approaches~\cite{liem_effective_2016} and some authors have built UV-complete theories in more specific frameworks, such as supersymmetric theories, describing the dark matter particles in our universe. More recently, some models known as simplified dark matter models have been proposed, in which the minimal particle content and only very few parameters are added in order to respect the different constraints set on DM properties from collider experiments and astrophysical observations, and ensuring a dark matter candidate stable enough to account for his abundance. In this kind of setup, several cases have been studied in which the DM particles can interact with SM particles through the exchange of a particle such as Higgs portal or Z portal~\cite{arcadi_z-portal_2015}\cite{arcadi_invisible_2014}. The framework of this project lies between a simple dark matter model and a UV complete Grand Unification Theory (GUT) description embedding the standard model and a dark matter candidate. We study the phenomenological aspects and derive constraints on DM properties in several scenarios considering interactions between a fermionic dark matter candidate and SM fields via the exchange of a massive spin-1 $Z^\prime$ in GUT inspired constructions.

\section{$Z^\prime$ portal scenarios in GUT inspired constructions}

In the context of GUT, a massive $Z^\prime$ can arise from the breaking of the GUT symmetry group at some intermediate scale~\cite{langacker_physics_2009}. In this work, we are considering two concrete realizations based on the $E_6$ group, namely $\chi$ and $\psi$ and a string inspired scenario, namely $\eta$ corresponding to the linear combination : $Z_\eta = \sqrt{\frac{3}{8}}Z_\chi-\sqrt{\frac{5}{8}}Z_\psi$. We consider as well the left-right symmetric model as an intermediate scale gauge group and the B-L model, both based on the $SO(10)$ GUT gauge group, and the Sequential Standard Model (SSM) as a reference. In this framework, we can parametrize the relevant part of the lagrangian in the following way :

\begin{equation}
\mathcal{L} \supset \tilde{g} \sum_f \bar{f}\gamma^{\mu}(\epsilon_L^f P_L + \epsilon_R^f P_R)f Z^\prime_{\mu}+g_{\chi}\bar{\chi}\gamma^{\mu}(\epsilon_L^\chi P_L + \epsilon_R^\chi P_R)\chi Z^\prime_{\mu}
\end{equation}

Where $f$ denotes the SM fermions, $\chi$ the DM particles, $\tilde{g}$ and $g_\chi$ denote gauge couplings and $\epsilon_{L,R}$ the couplings of the left and right handed components of the DM and SM fields.
We further assume that the specific representation of the GUT group in which the DM lies is unknown and we study the phenomenological aspects in a general case, as a result we consider the DM mass and its couplings as free parameters.
The SM fermions couplings $\epsilon^f_{L,R}$ to $Z^\prime$ are fixed by GUT construction and specified in Table~\ref{tab:couplings}, where $D$ is a normalization factor. We can re-express the lagrangian in a more convenient "$V-A$" parametrization :

\begin{equation}
\mathcal{L} \supset \tilde{g} \left( \sum_f \bar{f}\gamma^{\mu}(V_{f}-A_{f}\gamma^{5})f Z^\prime_{\mu}+\bar{\chi}\gamma^{\mu}(V_{\chi}-A_{\chi}\gamma^{5})\chi Z^\prime_{\mu} \right)
\end{equation}
\noindent
where : 
\begin{align}
&\tilde{g} V_{f}=\frac{\tilde{g}}{2D}(\epsilon_{L}^{f}+\epsilon_{R}^{f}), \qquad \tilde{g} A_{f}=\frac{\tilde{g}}{2D}(\epsilon_{L}^{f}-\epsilon_{R}^{f})\\
&\tilde{g} V_{\chi}=\frac{g_\chi}{2D}(\epsilon_{L}^{\chi}+\epsilon_{R}^{\chi}), \qquad \tilde{g} A_{\chi}=\frac{g_\chi}{2D}(\epsilon_{L}^{\chi}-\epsilon_{R}^{\chi})
\end{align}
\noindent
In the case of the SSM, we assumed $\tilde{g}=g \approx 0.65$ and in the GUT models, we choose the GUT normalized value $\tilde{g}=g_\chi=\sqrt{\frac{5}{3}} g \tan \theta_{W} \approx 0.46$ where $g$ denotes the gauge coupling of the SM $U(1)_Y$ gauge group.

\begin{table}
\begin{center}

\begin{tabular}{|c|c|c|c|c|c|c|}
\hline
 &$\chi$&$\psi$&$\eta$&LR&B-L&SSM\\
\hline
D&2$\sqrt{10}$&2$\sqrt{6}$&2$\sqrt{15}$&$\sqrt{5/3}$&1&1\\
\hline
$\epsilon^{u}_{L}$&-1&1&-2&-0.109&1/6&$\frac{1}{2}-\frac{2}{3}\sin^{2}(\theta_{W})$\\
\hline
$\epsilon^{d}_{L}$&-1&1&-2&-0.109&1/6&$-\frac{1}{2}+\frac{1}{3}\sin^{2}(\theta_{W})$\\
\hline
$\epsilon^{u}_{R}$&1&-1&2&0.656&1/6&$-\frac{2}{3}\sin^{2}(\theta_{W})$\\
\hline
$\epsilon^{d}_{R}$&-3&-1&-1&-0.874&1/6&$\frac{1}{3}\sin^{2}(\theta_{W})$\\
\hline
$\epsilon^{\nu}_{L}$&3&1&1&0.327&-1/2&$\frac{1}{2}$\\
\hline
$\epsilon^{l}_{L}$&3&1&1&0.327&-1/2&$-\frac{1}{2}+\sin^{2}(\theta_{W})$\\
\hline
$\epsilon^{l}_{R}$&1&-1&2&-0.438&-1/2&$\sin^{2}(\theta_{W})$\\
\hline
\end{tabular}
\end{center}
\caption{Left and right handed SM fermion couplings to $Z^\prime$ and GUT normalization factor $D$ for the several models considered in this work}
\label{tab:couplings}

\end{table}

\section{Confronting relic density with direct detection constraints}
In this section we investigate the Spin Dependent (SD) and Spin Independent (SI) direct detection phenomenological constraints by deriving general results in the several models considered in this work and we confront the canonical thermally produced DM scenario with direct detection exclusion limits in these GUT inspired scenarios.

The computation of the SI and SD scattering cross section between DM particles and nucleons is straightforward and can be expressed, for a proton as an example, in the following way :

\begin{equation}
\sigma_{SI}^{p}=\frac{\mu_{\chi p}^{2}\tilde{g}^{4}V_{\chi}^{2}}{\pi M_{Z^{\prime}}^{4}}\alpha_{SI}, \qquad \qquad \sigma_{SD}^{p}=\frac{3\mu_{\chi p}^{2}\tilde{g}^{4}A_{\chi}^{2}}{\pi M_{Z^{\prime}}^{4}}\alpha_{SD}
\label{sigmasisd}
\end{equation}

Where $\mu_{\chi p}$ is the DM-proton reduced mass, $M_{Z^\prime}$ is the $Z^\prime$ mass. Notice here that the SI contribution of the scattering cross section will only be dependent on the vectorial coupling $V_\chi$ to DM and on the other hand, the SD contribution will be sensitive to $A_\chi$ only. \\
$\alpha_{SI}$ and $\alpha_{SD}$ are correction factors defined to take into account the non identical DM-neutron and DM-proton interaction, which is commonly assumed by experimental collaborations when extracting constraints on scattering cross section, and taking into account the nuclear isotopes abundances of the detector material :
\begin{subequations}
\begin{equation}
\alpha_{SD}=\frac{\sum_{A}\eta_{A}A^2[V_{u}(1+Z/A)+V_{d}(2-Z/A)]^{2}}{\sum_{A}\eta_{A}A^{2}}
\end{equation}
\begin{equation}
\alpha_{SI}=\frac{\sum_{A}\eta_{A}[A_{u}(\Delta^{p}_{u}S^{A}_{p}+\Delta^{n}_{u}S^{A}_{n})+A_{d}(\Delta^{p}_{d}S^{A}_{p}+\Delta^{n}_{d}S^{A}_{n}+\Delta^{p}_{s}S^{A}_{p}+\Delta^{n}_{s}S^{A}_{n})]^{2}}{\sum_{A}\eta_{A}[S^{A}_{p}+S^{A}_{n}]^{2}}
\end{equation}
\end{subequations}

Where $Z$ ($A$) is the number of proton (nucleons) of each isotope and $\eta_A$ is the relative abundance of the isotope with $A$ nucleons. The coefficients $\Delta_q^N$ denote the contribution of the quark $q$ to the spin of the nucleon $N$ and $S_N^A$ the contribution of the nucleon $N$ to the spin of a nucleus with $A$ nucleons.

The ratio of the SD contribution of the cross section over the SI contribution can be expressed as a function of the parameter $\alpha=A_\chi/V_\chi$ in the following way :
\begin{equation}
\frac{\sigma_{SD}^{p}}{\sigma_{SI}^{p}}=3\alpha^{2}\frac{\alpha_{SD}}{\alpha_{SI}}, \qquad \frac{\sigma_{SD}^{n}}{\sigma_{SI}^{n}}=3\alpha^{2}\frac{\alpha_{SD}}{\alpha_{SI}}\left(\frac{2V_{d}+V_{u}}{2V_{u}+V_{d}}\right)^{2}\left(\frac{A_{u}\Delta_{u}^{p}+A_{d}(\Delta_{d}^{p}+\Delta_{s}^{p})}{A_{u}\Delta_{u}^{n}+A_{d}(\Delta_{d}^{n}+\Delta_{s}^{n})}\right)^{2}
\label{eq:ratiosdsi}
\end{equation}

\begin{center}
\begin{figure}[!h]
  \begin{minipage}[c]{0.5\textwidth}
  \includegraphics[width=0.9\textwidth]{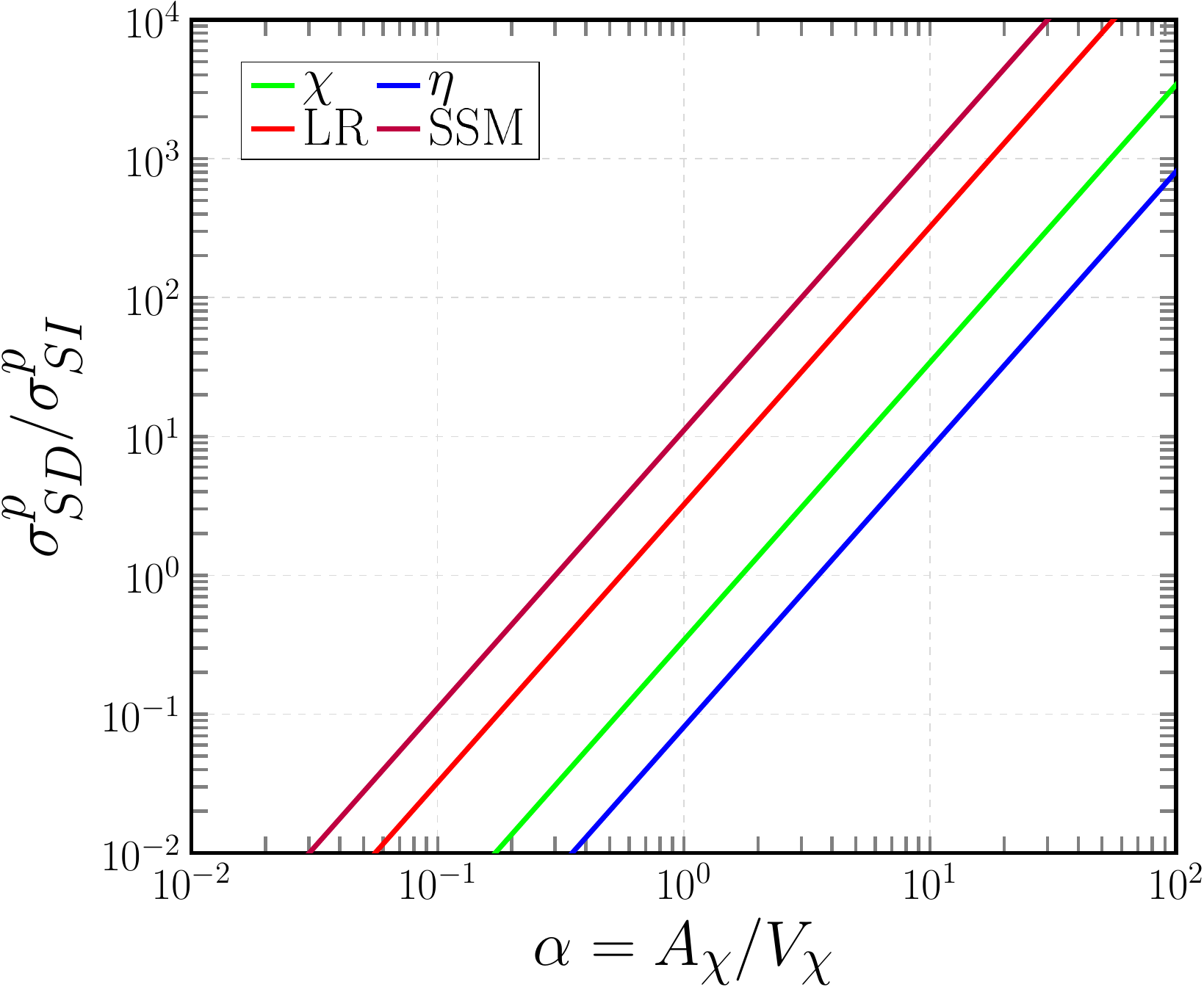}
   \end{minipage}\hfill
   \begin{minipage}[c]{0.5\textwidth}   
      \includegraphics[width=0.9\textwidth]{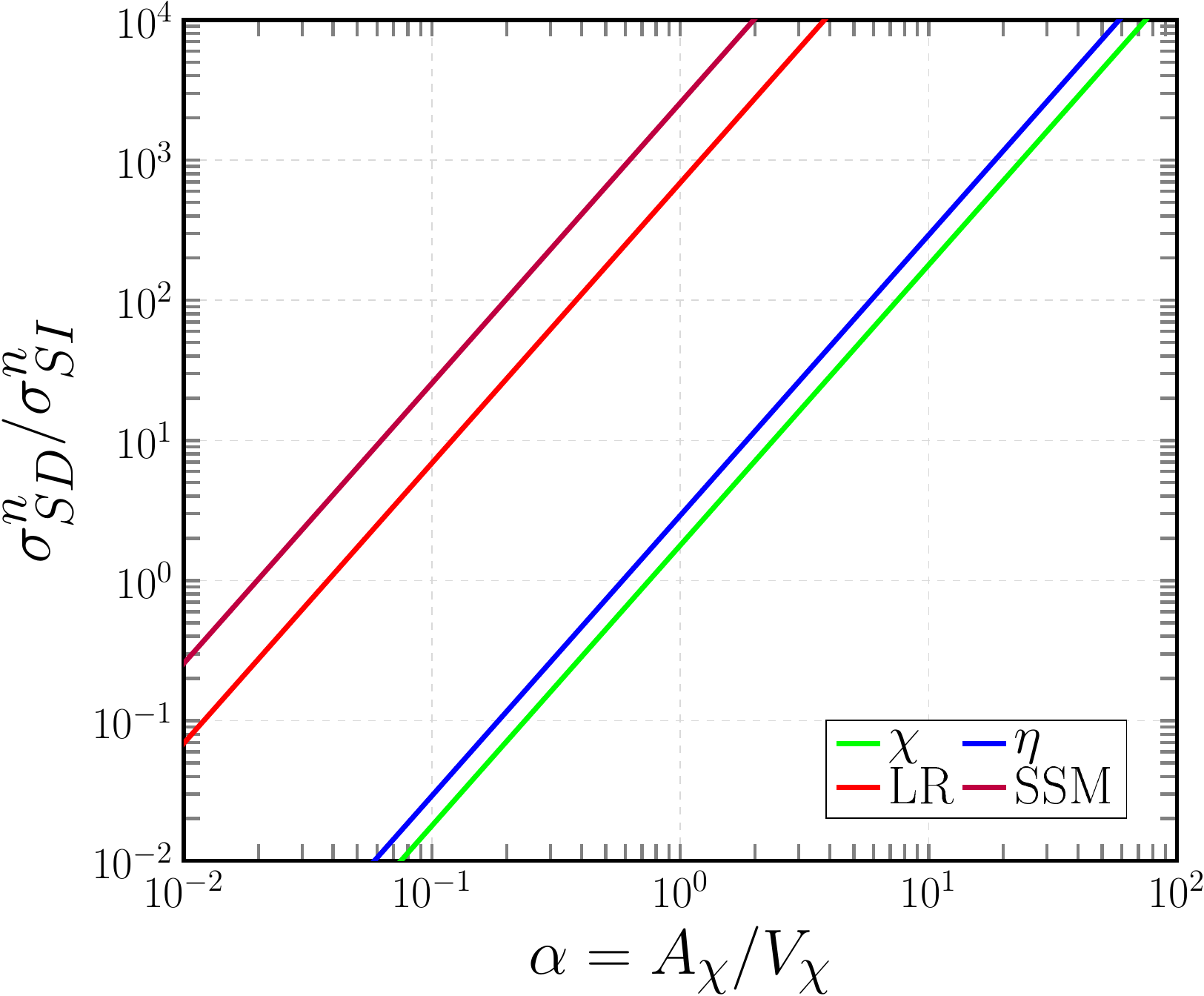}
   \end{minipage}
   \caption{Evolution of the ratio of the SD over the SI dependent contribution of the cross section for the proton (left) and neutron (right) with the nature of the dark matter coupling to $Z^\prime$, parametrized by the parameter $\alpha$}
	\label{fig:sisd}
\end{figure}
\end{center}

Notice that these relations are valid as long as the dark matter mass is of the order of the GeV scale and the $Z^\prime$ mass is much larger, but they do not rely on any other specific assumption and are valid for each model considered in this work. The evolution of $\frac{\sigma_{SD}}{\sigma_{SI}}$ for protons and neutrons with the parameter $\alpha$ has been represented in Figure~\ref{fig:sisd} for the scenarios considered in this work. In the case of the B-L scenario, the right and left handed couplings are identical implying a purely vectorial coupling between quarks and $Z^\prime$, thus not presented on this plot as well as the $\psi$ scenario, predicting only axial couplings.

In the canonical thermally produced DM scenario, the DM particles decoupled from the SM thermal bath while being non relativistic with a usually assumed maxwellian velocity distribution centered around $v \sim 0.3$. The relic density is related to the velocity averaged annihilation cross section $<\sigma v>$ via the Boltzmann equation and the correct relic abundance can be reached for $<\sigma v>\simeq 3.10^{-26}\text{cm}^3\text{s}^{-1}$ for a DM mass of $O$(GeV). A reliable analytical estimate of $<\sigma v>$ is obtained by performing a velocity expansion~\cite{gondolo_cosmic_1991} :
\begin{equation}
<\sigma v>\approx \frac{\tilde{g}^4m_\chi^2}{\pi M_{Z^\prime}^4}\sum_f c_f (A_f^2+V_f^2)V_\chi^2\left(1+\frac{2}{3}(2\alpha^2-1)v^2 \right)
\end{equation}
\begin{center}
\begin{figure}[!h]
  \begin{minipage}[c]{0.5\textwidth}
\includegraphics[width=0.95\textwidth]{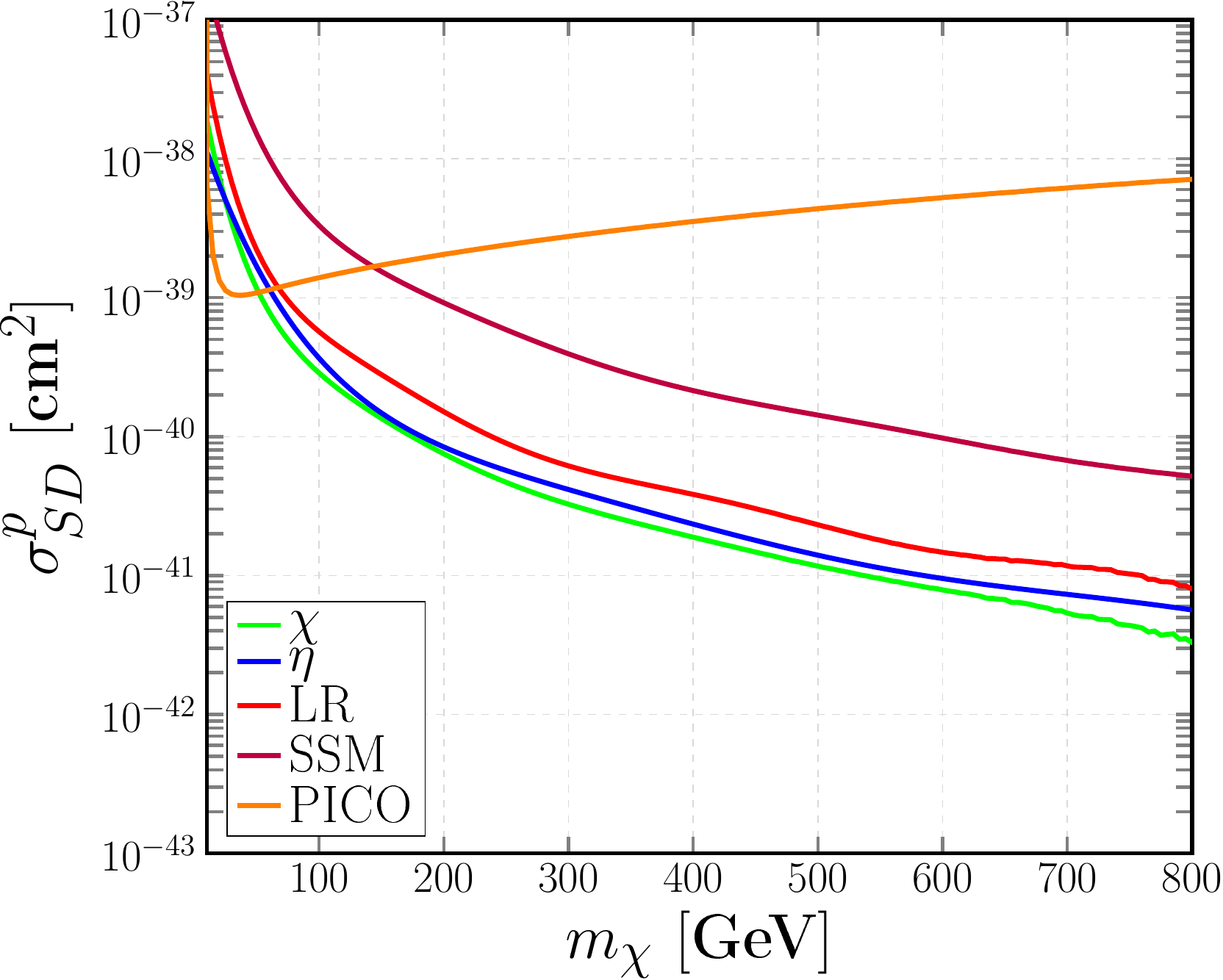}   \end{minipage}\hfill
   \begin{minipage}[c]{0.5\textwidth}   
\includegraphics[width=0.95\textwidth]{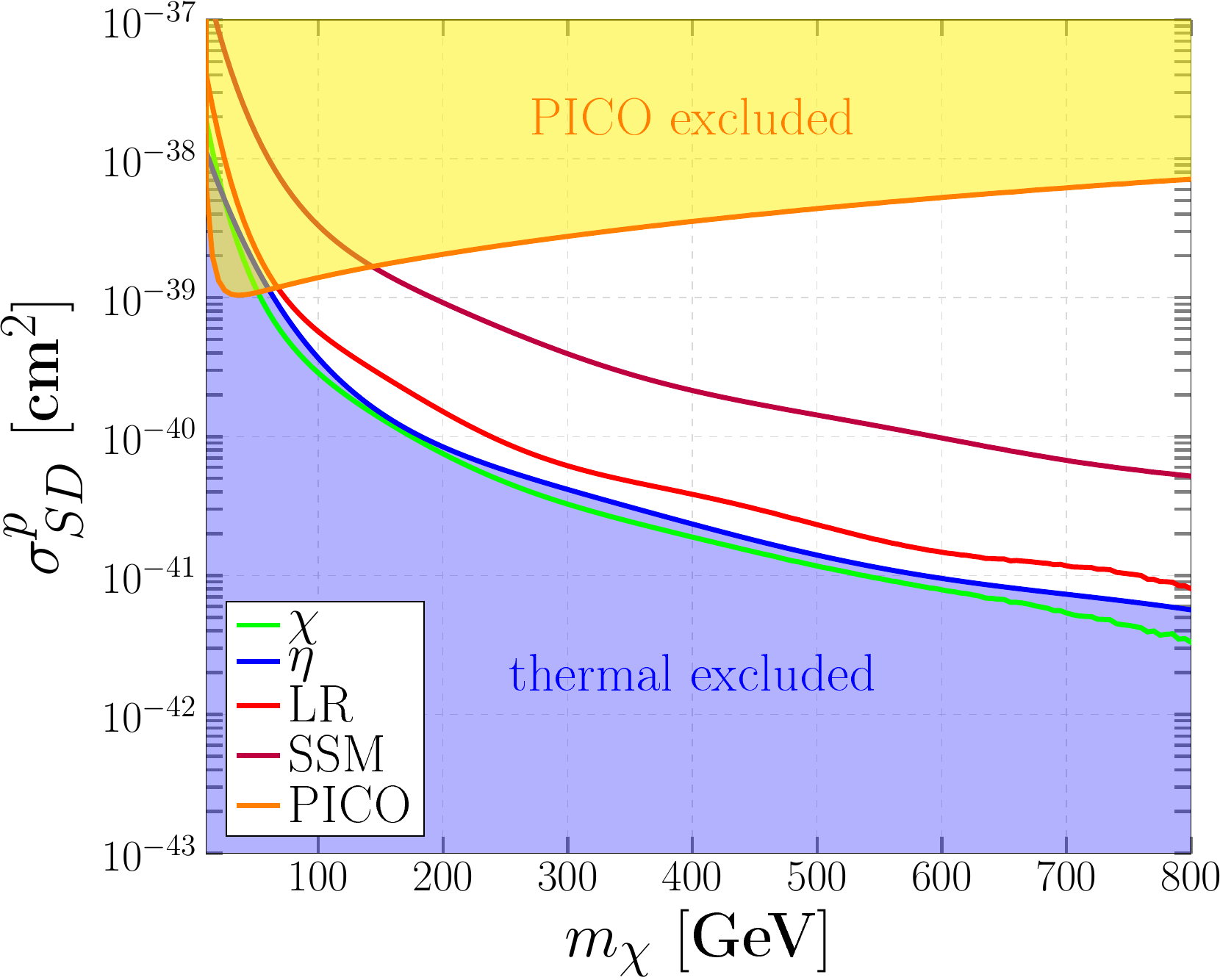}   \end{minipage}
   \caption{On the left, $\sigma^p_{SD}$ as a function of the DM mass for the $Z^\prime$ theories considered, assuming a DM relic density matching the Planck observations and considering the limits on $\sigma^p_{SI}$ derived by the LUX collaboration. The constraints on $\sigma^p_{SD}$ deduced from the measurements of the PICO collaboration are shown in orange. On the right, we show the parameter space excluded by PICO and excluding a thermally produced DM scenario in the case of the $\eta$ model.}
	\label{fig:sigmasdlux}
\end{figure}
\end{center}

Where $c_f$ is a color factor. Here we neglected higher orders in $v^2$ and we neglected the masses of the SM fermions. This hypothesis remains valid for $m_\chi$ $\gtrsim$ 10 GeV for most of the SM fermions except for the top quark\footnote{This hypothesis is made only for illustrative purposes, the results are almost unaffected by considering its contribution.}. Notice here that in this expression the dominant contribution will be triggered by the vectorial coupling, and therefore the SI contribution of the direct detection scattering cross section from equation~\eqref{sigmasisd}. In this framework, the velocity averaged annihilation cross section depends only on four parameters : the DM mass, the $Z^\prime$ mass, the vectorial coupling $V_\chi$ and the parameter $\alpha$. The LUX collaboration has set strong constraints on the SI cross section~\cite{lux_collaboration_improved_2015} therefore on the vectorial DM coupling. Respecting these constraints and assuming that the DM abundance fits the Planck observations, there is a straightforward relation between the DM and $Z^\prime$ masses and the SD cross section through equation~\eqref{eq:ratiosdsi}. Therefore we can reinterpret the constraints set by LUX on SI cross section into SD constraints, as a function of the dark matter mass. In principle the results are dependent on $M_{Z^\prime}$ but as long as the condition $M_{Z^\prime}\gg m_\chi$ is verified, by varying $M_{Z^\prime}$ the results are almost unchanged and thus our results remain general. The derived constraints on the SD cross section are presented Figure~\ref{fig:sigmasdlux} for each model considered, where we show the current exclusion limits on $\sigma^p_{SD}$ set by the PICO collaboration~\cite{amole_dark_2015} as well. For each model, the only viable parameter space lies between the several curves and the PICO exclusion limits, as represented on Figure~\ref{fig:sigmasdlux} on the right. Indeed, assuming a very low SD cross section, the axial coupling $A_\chi$ would be suppressed with respect to $V_\chi$, leading the DM density to be only dependent on the vectorial coupling $V_\chi$. In order to avoid being in tension with the LUX exclusion limits, the coupling $V_\chi$ will have to be small leading to an overabundant DM. The intermediate values of $\sigma^p_{SD}$ allow the DM density to depend significantly on $A_\chi$ and avoid having a large $V_\chi$ giving the correct density, leading to an under abundant DM but still a viable scenario. At DM masses $\lesssim 100$ GeV, all of the different scenarios considered here are in tension with a thermally produced dark matter scenario and the exclusion limits set by the PICO collaboration.\\

\section{Conclusion and prospects}

In this work we showed that by embedding simplified dark matter models in a general $E_6$ and $SO(10)$ grand unification theory constructions, we can combine and use the complementary of spin independent and spin dependent constraints from direct detection with LUX and PICO data, and observations from Planck, to derive stringent bounds on the thermally produced dark matter scenario and exclude dark matter masses below $\sim 100$ GeV in several different models in this setup without specific assumptions on the dark matter couplings. Collider analysis have been performed in general extensions of the standard model involving a $Z^\prime$, and combining cosmological measurements and constraints from the upcoming direct detection experiments and results from run II of the LHC will be a powerful tool to probe a larger parameter space in our framework and might lead to a reconsideration of the thermally produced dark matter paradigm.

\bibliographystyle{unsrt}

\bibliography{biblio}

\end{document}